\documentclass[a4paper,12pt]{article}
\usepackage{amssymb}
\usepackage{amsmath}
\usepackage{graphicx}
\usepackage{multicol}
\usepackage{bm}
\usepackage{epstopdf,lineno,hyperref}
\usepackage{color}
\usepackage[toc,page]{appendix}
\usepackage{amsmath}
\usepackage{amsthm}
\usepackage{amssymb}
\usepackage{graphicx}
\usepackage{multirow}
\usepackage{setspace}
\usepackage{url}
\usepackage{subfigure}
\usepackage{bbm}
\newcommand*{\Scale}[2][4]{\scalebox{#1}{$#2$}}%

\begin{document}


\title{Multiverse effects on the CMB angular correlation function in the framework of NCG}


\author{Sahar Arabzadeh $^{1}$ \footnote{sr.arabzadeh@gmail.com} and Kamran Kaviani $^{2}$\footnote{kamran.kaviani@gmail.com} \\
\footnotesize {$^{1,2}$ Department of Physics, Alzahra University, Tehran, Iran} \\}
\date{}
\maketitle

\begin{abstract}
Following many theories that predict the existence of the multiverse and by conjecture that our space-time may have a generalized geometrical structure at the fundamental level, we are interested in using a non-commutative geometry (NCG) formalism to study a suggested two-layer space that contains our 4-dimensional (4-D) universe and a re-derived photon propagator. It can be shown that the photon propagator and a cosmic microwave background (CMB) angular correlation function are comparable, and if there exists such a multiverse system, the distance between the two layers can be estimated to be within the observable universe's radius. Furthermore, this study revealed that our results are not limited to CMB but can be applied to many other types of radiation, such as X-rays. 
\end{abstract}



\section{Introduction}

There are many theories about the early universe that predict the existence of the multiverse. The concept of such a system was first proposed in the many-world interpretation of quantum mechanics. Since then, it has been studied within the framework of string theory\cite{Everett}, so the idea of an inflationary multiverse is derived from the idea of eternal inflation \cite{Linde, Garica}. In the standard picture, there exist infinite number of pocket universes (bubbles) such that each universe corresponds to a vacuum. Each pocket universe contains an infinite number of universes; in that case, each pocket universe has a positive cosmological constant \cite{Bousso}. However, the multiverse comprises a larger physical structure which contains our 4-D universe as well. The other universes may lie beyond our observable universe: by that definition, they are unobservable. Therefore, it is imperative that researchers find observable evidence of the existence of the multiverse. There have been various researches conducted on a cosmic microwave background (CMB) that may provide that observable evidence \cite{Salem, Kaleban}. In this manuscript, we examine the effect of the multiverse on the two-point function of quantum electrodynamics (QED) in a two-layer space within a framework of non-commutative geometry (NCG). We chose this framework following Connes' analogue, according to the conjecture that our space-time may have a generalized geometrical structure at a fundamental level.

This manuscript is organized into several sections. In section \ref{ms}, we start with the mathematical set-up which is sufficient for our present job. In the next section, a two-layer space is examined as a multiverse, and we investigate the effect of such a system on a photon propagator of our 4-D universe. The resulting potential can be compared with a CMB two-point angular correlation function, as the observable evidence of the existence of a multiverse. The results showed that the Coulomb potential in each layer of this two-layer system and the CMB two-point angular correlation are comparable. Furthermore, if there exists such a multiverse system, a distance of two layers can be estimated to be within the observable universe's radius. In addition, the corresponding Coulomb potential eventuates a tiny redshift in hydrogen atom levels, so that if the distance between two layers is considered to be within the observable universe's radius, the mentioned redshift will be on the order of $10^{-25} eV$. We will also show that this result is not limited to CMB but can be applied to all monochromatic waves, such as X-rays.

\section{Mathematical setup}
\label{ms}

In NCG, a spectral triple ($A, H, D$) includes all geometrical information of a non-commutative space in which $A$ is an involutive algebra of bounded operators on a Hilbert space $H$, and the generalized Dirac operator $D$ is a self-adjoint operator on $H$ but does not belong to $A$ \cite{Connes}. As an example for ordinary manifolds, the algebra $A$ can be the algebra of smooth functions on manifolds. The Hilbert space is the vector space of Square-integrable functions. This method can be generalized to the other $C^*$ algebras which are not commutative and so there is no corresponding manifold for them. 

In this approach, The Dirac operator plays a metric role and determines the distance between two points (states) of the geometrical space, using the Connes' formula of distance \cite{Connes, Landi, Connes 2}:
  
\begin{equation}
d(1,2)=\sup_{a\in A} \lbrace \vert a_1 - a_2 \vert : \Vert [D,a] \Vert \leq 1\rbrace .
\end{equation}

In the above formula $a_1$ and $a_2$ are elements of the algebra. One can easily obtain and re-derive usual distance in a commutative space by using the above formula. In many cases the term $[D,a]$ denotes by $d a$ means differential of $a$. We remember that the commutator satisfies Leibniz rule. In addition, a Connes' \textit{p-form}, $\omega_p$ can be derived from the following relation \cite{Landi}:

\begin{equation}
\label{omega}
\omega_p=\sum\limits_j a_0^j [D,a_1^j] [D,a_2^j]\;...\;[D,a_p^j]\;\;\; , a_i^j\in A,
\end{equation}

In NCG fiber bundles can be presented as modules over algebra. By using Serre-Swan theorem \cite{Connes, Landi}, one may generalize the Yang-Mills field which is the connection on fiber bundle by introducing a self-adjoint algebraic one-form as the gauge field. Suppose $\mathbb{A}$ exists as a self-adjoint one-form that represents a gauge potential. The corresponding field strength or curvature, $\theta$, is defined by $\theta = \mathbb{A}^2 + d\mathbb{A}$, which is a two-form, and the Yang-Mills action function is of the form:

\begin{equation}
\label{yms}
\mathcal{S}= YM(\mathbb{A})\equiv tr_\omega ((\mathbb{A}^2 + d\mathbb{A})^2),
\end{equation}
where $tr_\omega$ is the Dixmier trace \cite{Landi} and plays the role of ordinary trace and the role of integration for discrete part of space and over the continues structures respectively.

The following equation shows the relation between the two-layer space action functional and corresponding Lagrangian.

\begin{equation}
    \mathcal{S}= \int tr ((\mathbb{A}^2 + d\mathbb{A})^2)  d^4 x
\end{equation}

Note that the discrete dimension of the two-layer space is zero so we just drop the $\int d^4 x$ to write the Lagrangian \cite{Landi}. 

\begin{equation}
\label{lagra}
    \mathcal{L}= tr ((\mathbb{A}^2 + d\mathbb{A})^2)
\end{equation}

\section{Two-layer space}
\label{tls}

The underlying assumption is that, within a multiverse, separated universes have only gravitational interactions. Suppose a set of points. A constraint on some subset of these points to have no Yang-Mills interaction with others creates \textit{a partition of the set}. One may perform this by defining an equivalence relation as: point $i\sim$ point$j$, if only there exists a Yang-Mills interaction between them. Let us call each equivalence class a universe (see FIG. 1). These universes have only gravitational interactions with each other. In terms of NCG, they have only geometrical interactions. It is expected after these considerations that the distance between each two universes should be so large as the long-range interactions can be ignored.

\begin{figure}[h]
\centering
\includegraphics[width=5.5cm]{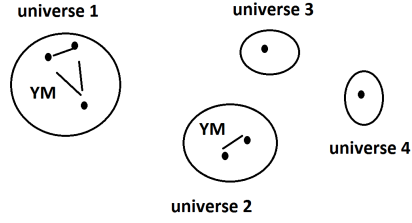}
\caption{We call each subset of points that have Yang-Mills interaction with each other,a universe.}
\end{figure}


We assume that there exists a universe like ours at a specific distance. That universe is studied in the framework of NCG by consideration of a two-layer space, each layer containing a 4-D compact spin manifold denoted by $Y$. To introduce the desired spectral triple for this space, the algebra $A$ is the algebra of block diagonal matrices in which their entries are smooth real functions, $A= C_\mathbb{R}^\infty (Y) \oplus C_\mathbb{R}^\infty (Y)$. The Hilbert space $H$ is the direct sum of two Hilbert spaces on $Y$, $H= H(Y) \oplus H(Y)$.  

The entries of the Dirac operators are first-order differential operators that act on the spinors of $Y$. Due to the existence of $\mathbb{Z}_2$ equivalence, the block diagonal entries of $D$ and the off-diagonal elements are equal, respectively. In addition, because $[D, a]$ for $a \in A$ is a multiplication operator, off-diagonal blocks should be a multiplication operator \cite{Chams}, which results in a self-adjoint operator. Therefore, the most general form of the Dirac operator is as follows:

\begin{equation}
\label{dirac}
D= \begin{pmatrix}
i \gamma_\mu (\partial^\mu+ ie\mathbb{A}^\mu+\cdots) & \psi+ \gamma^5 m\\
\psi+ \gamma^5 m & i \gamma_\nu (\partial^\nu+ ie\mathbb{A}^\nu+\cdots)\\
\end{pmatrix},
\end{equation}

where $\psi$ and $m$ are real functions. $\mathbb{A}$ is the gauge potential. Suppose that $\psi$ is equal to unit, and we limit our calculations to QED for simplicity.

By using equation (2), the generalized one-form is as follows:

\begin{equation}
\label{phi}
\omega=\begin{pmatrix}
\gamma^\mu \omega_{1 \mu} & \bar{\gamma} \phi\\
-\bar{\gamma} \phi^{\dagger} & \gamma^\nu \omega_{2 \nu}\\
\end{pmatrix}.
\end{equation}

where $\bar{\gamma}=\psi+\gamma^5 m$. Using the above $\omega$ as a Yang-Mills field, one may write the field strength, $\theta$ as:

\begin{equation}
\label{scale}
\Scale[0.78]{
\theta=\begin{pmatrix}
\gamma^{\mu\nu}(\partial_\mu \omega_{1 \nu}-\partial_\nu \omega_{1 \mu})+2m\gamma^5(\phi-\phi^{\dagger})-\bar{\gamma}^2(\phi^{\dagger}\phi)  & (m+\phi) \gamma^\mu \gamma^5 (\omega_{1 \mu} - \omega_{2 \mu})- m \gamma^{5} \gamma^{\mu} D_{\mu} \phi^\dagger \\
-(m+\phi) \gamma^\mu \gamma^5 (\omega_{1 \mu} - \omega_{2 \mu})+ m \gamma^{5} \gamma^{\mu} D_{\mu} \phi &\gamma^{\mu\nu}(\partial_\mu \omega_{2 \nu}-\partial_\nu \omega_{2 \mu})+2 m \gamma^5 (\phi-\phi^{\dagger})-\bar{\gamma}^2(\phi^{\dagger}\phi) \\
\end{pmatrix},}
\end{equation}

where $\gamma^{\mu\nu} =\frac{1}{2}(\gamma^\mu \gamma^\nu - \gamma^\nu \gamma^\mu) $. See Appendix A for more calculations. The one-form, $\omega$, and curvature, $\theta$, are consistent with Chamseiddine et al.'s results for the two-layer space \cite{Chams}. By substituting $e\mathbb{A}_{i\mu}$ as the $\omega_{i \mu}$ and by using equation (\ref{lagra}), one can easily write the Yang-Mills Lagrangian as follows:

\begin{equation}
\label{Lagr}
\begin{array}{c}
\mathcal{L}=-8e(( \partial^\mu \mathbb{A}^\nu_1-\partial^\nu \mathbb{A}_1^\mu)(\partial_\mu \mathbb{A}_{1 \nu}-\partial_\nu \mathbb{A}_{1\mu})+\\
\\
( \partial^\rho \mathbb{A}_2^\sigma-\partial^\sigma \mathbb{A}_2^\rho)(\partial_\rho \mathbb{A}_{2\sigma}-\partial_\sigma \mathbb{A}_{2\rho})-(m+\phi)^2(\mathbb{A}_1^\mu -\mathbb{A}_2^\mu)(\mathbb{A}_{1 \mu}-\mathbb{A}_{2 \mu}))\\
\\
- 4 m^2 D_\mu \phi^\dagger D^\mu \phi + 2(1+m^2)^2 (\phi^\dagger \phi)^2 -8m^2 (\phi^\dagger \phi)\\
\end{array},
\end{equation}

Let us return to equation (\ref{phi}) and shield the Yang-Mills field. Then, having two almost non-Yang-Mills interacting 4-D universes by tuning $\phi$, $\phi$ connects the two layers through Yang-Mills interactions. To retain the gauge invariance, we keep the field $\phi$. However, we take it as tiny as possible to be able to continue shielding the Yang-Mills field. The dominant terms of the Lagrangian function become the following form:

\begin{equation}
\label{Lag}
\begin{array}{c}
\mathcal{L}=-8e(( \partial^\mu \mathbb{A}^\nu_1-\partial^\nu \mathbb{A}_1^\mu)(\partial_\mu \mathbb{A}_{1 \nu}-\partial_\nu \mathbb{A}_{1\mu})+\\
\\
( \partial^\rho \mathbb{A}_2^\sigma-\partial^\sigma \mathbb{A}_2^\rho)(\partial_\rho \mathbb{A}_{2\sigma}-\partial_\sigma \mathbb{A}_{2\rho})-\\
\\
 m^2(\mathbb{A}_1^\mu -\mathbb{A}_2^\mu)(\mathbb{A}_{1 \mu}-\mathbb{A}_{2 \mu}))\\
\end{array}.
\end{equation}

By using the above Lagrangian and choosing a Feynman gauge, one may obtain a photon propagator as follows (see Appendix B for more details):

\begin{equation}
\label{prop}
\tilde{D}_F^{\mu\nu}(k) = \frac{i m^2g^{\mu\nu}}{k^2(k^2+2m^2) - i \varepsilon}.
\end{equation}

This propagator has two poles, which suggests a mass-less photon and a tachyon-type photon with a mass order of $m$. Since we do not see evidence of such photons (for example, in the thermodynamic properties), we expect $m$ to be small enough.

We will study two consequences of choosing this propagator in the subsections \ref{scf} and \ref{cp}.

\subsection{The spatial correlation function}
\label{scf}

We are now ready to compare the resultant propagator with the two-point angular correlation function from the Wilkinson Microwave Anisotropy Probe (WMAP) nine-year results \cite{Copi, CMB}. The spatial correlation function can be calculated from the Fourier transform of equation (\ref{prop}) and then integrated on the time because within the CMB, we detect photons from all time ranges in each specific direction.

\begin{equation}
\label{new}
\begin{array}{c}
\tilde{D}_F^{\mu\nu}(r)\equiv \int_0^T \mathrm{d} t \tilde{D}_F^{\mu\nu}(x) = \int_0^T \mathrm{d} t \int \frac{\mathrm{d}^4 k}{(2\pi)^4}\;\frac{i m^2g^{\mu\nu} \mathrm{e}^{-ikx}}{k^2(k^2+2m^2) - i \varepsilon} \\
\\
= \int_0^T  \mathrm{d} t \int_{-\infty}^{+\infty} \mathrm{d}k_0 \int \frac{ \mathrm{d}^3 k}{(2\pi)^4}\;\frac{i m^2g^{\mu\nu}}{k^2(k^2+2m^2) - i \varepsilon} \mathrm{e}^{-ik_0 t}\mathrm{e}^{i\vec{k}.\vec{r}}=\\
\\
\int_{-T}^T \mathrm{d} t \int_{0}^{+\infty} \mathrm{d}k_0 \int \frac{ \mathrm{d}^3 k}{(2\pi)^4}\;\frac{i m^2g^{\mu\nu}}{k^2(k^2+2m^2) - i \varepsilon} \mathrm{e}^{-ik_0 t}\mathrm{e}^{i\vec{k}.\vec{r}}= \\
\\
\int_{0}^{+\infty} \mathrm{d}k_0 \int \frac{ \mathrm{d}^3 k}{(2\pi)^4}\;\frac{2 i m^2g^{\mu\nu} \mathrm{e}^{i\vec{k}.\vec{r}}}{(k_0^2 -\vert k \vert^2)(k_0^2 -\vert k \vert^2+2m^2)} \frac{sin(k_0 T)}{k_0}=\\
\\
\int \frac{\mathrm{d}^3 k}{(2\pi)^3}\;\frac{i m^2g^{\mu\nu}\mathrm{e}^{i\vec{k}.\vec{r}}}{\vert k \vert ^2(\vert k \vert^2- 2m^2) - i \varepsilon}+\int \frac{\mathrm{d}^3 k}{(2\pi)^3}\;\frac{i m^2g^{\mu\nu} cos(\vert k\vert T)}{\vert k \vert ^2(2m^2) - i \varepsilon}\\
\end{array}
\end{equation}

where $T$ is the age of the universe. The first term is a multiple of the Coulomb potential, as we will see in the next section. To have a better sense about the other term, consider a sinusoidal wave, which travels in limited-time intervals. Obviously, it is not a monochrome wave because its Fourier transform contains all frequency contributions. However, if it travels in the time from $-\infty$ to $+\infty$, it will be monochrome. Notice that in the case of the CMB correlation function, photons can be estimated to be monochrome (with a frequency of about 160.2 GHz). Now in the limit $T \to \infty$ and using the third line in equation (\ref{new}), one can easily show the following relation:

\begin{equation}
\begin{array}{c}
\tilde{D}_F^{\mu\nu}(r)=\int_{0}^{+\infty} \int \frac{\mathrm{d}k_0 \mathrm{d}^3 k}{(2\pi)^3}\;\frac{i m^2g^{\mu\nu}}{k^2(k^2+2m^2) - i \varepsilon} \mathrm{e}^{i\vec{k}.\vec{r}} \delta(k_0)\\
\\
=\int \frac{\mathrm{d}^3 k}{(2\pi)^3}\;\frac{i m^2g^{\mu\nu}}{\vert k \vert ^2(\vert k \vert^2- 2m^2) - i \varepsilon}\mathrm{e}^{i\vec{k}.\vec{r}}dt\\
\\
= \frac{i g^{\mu\nu}}{4 \pi}\left(\frac{cos(\sqrt{2}m r)}{r} \right)\\
\end{array}
\end{equation}

The $\delta(k_0)$ indicates that only low-frequency photons contribute to the propagator. This condition occurs when we study the correlation function of monochrome waves. Assuming that the radiation energy density at two separated points are almost the same, we expect that the energy of the propagating photon to be low.

\begin{figure}[h]
\centering
\includegraphics[width=7cm]{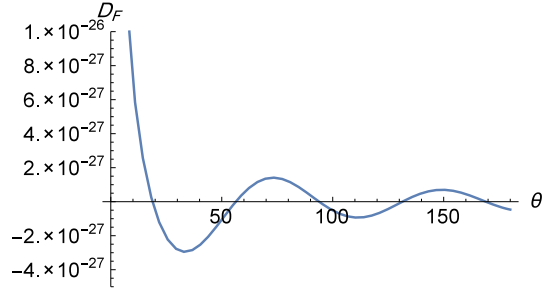}
\caption{Photon spatial correlation divided to $i\;g^{\mu\nu}$ vs. $\theta$.(for $m=0.03634 (G ly)^{-1}$)}
\end{figure}

\begin{figure}[h]
\centering
\includegraphics[width=7cm]{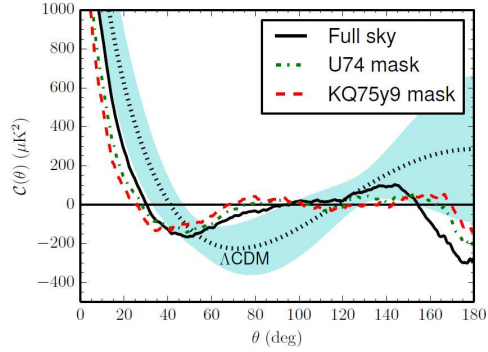}
\caption{The two-point angular correlation function from the WMAP 9 year results \cite{Copi}.}
\end{figure}

For better comparison with the CMB two-point correlation function, consider our universe effectively as a 1-dimensional ring with the radius of the observable universe $R$ (about 46 billion light years), where we are standing at the center of the so-called ring. We can substitute $r$ in the equation(\ref{vr}) with $R\theta$, where $\theta$ is a viewing angle. Since $m$ is dimensionally proportional to the inverse of the length, it can be substituted with $\frac{k}{R}$, where $k$ is some real factor. The FIG. 2 is plotted using $m=0.03634 (G ly)^{-1}$. This value is obtained by fitting the zeroes of two curves in FIG. 2 and the "KQ75y9 mask" in FIG .3. by using Least Mean Square method. The effect of increasing $m$ is increasing the oscillation of $D_F$. X-ray is a monochromatic wave too and one may see that the X-ray two-point angular correlation function has a similar behaviour \cite{Xray}. This is another observable effect of $m$ in our universe

Let us consider $C(\theta)$ as the angular two-point correlation function, which is defined as the average product between the temperature of two points' angles $\theta$ apart \cite{Sophie}.

\begin{equation}
C(\theta)=\overline{T(\hat{\Omega}_1),T(\hat{\Omega}_2)}\vert_{\hat{\Omega}_1.\hat{\Omega}_2=cos(\theta)}
\end{equation}

Here $T(\hat{\Omega})$ is the fluctuation around the mean value of the temperature, in direction $\hat{\Omega}$ in the sky. This temperature is microwave radiation, or  electromagnetic radiation, which is spatially averaged in all directions. Therefore, we can compare it with a QED spatial propagator.

FIG. 2 is plotted for $m=\frac{3.38}{R}$ in which a row estimation means that by considering $m$ as the inverse of distance between two layers, the distance between these two universes exists in the order of the radius of an observable universe. This is the observable evidence, as the signature of the multiverse within our 4-D universe.

\subsection{The effect on Coulomb potential and Hydrogen energy levels}
\label{cp}

In the case of two distinguishable fermions' scattering, the leading order contribution to the Feynman amplitude is an equation (\ref{lead}), where $p$, $p'$, $k$ and $k'$ are incoming and outgoing momentums respectively:

\begin{equation}
\label{lead}
\begin{array}{c}
i \mathcal{M}=
 - i e^2 \overline{u}(p') \gamma_\mu u(p) \frac{ g^{\mu\nu}  m^2}{q^2\left( q^2 +2m^2\right)-i \varepsilon}\overline{u}(k')\gamma_\nu  u(k),\\
\\
q=p'-p=k'-k.\\
\end{array}
\end{equation}

In the non-relativistic limit:

\begin{equation}
i \mathcal{M}= \frac{- i e^2 m^2}{\vert q\vert^2(\vert q\vert^2-2m^2)} (2m \eta '^\dagger \eta) (2m \eta '^\dagger \eta),
\end{equation}

where $\eta$ is a two-component constant spinor. We compare this with the Born approximation for the scattering amplitude within non-relativistic quantum mechanics:

\begin{equation}
\langle p'\vert iT\vert p\rangle = -i\tilde{V}(q) (2\pi)\delta (E_{p'}- E_p).
\end{equation}

For the QED interaction in this kind of multiverse system, the Coulomb potential in the momentum and coordinate space are in the following forms, respectively (See Appendix C for more details.):

\begin{equation}
\label{vr}
\begin{array}{c}
\tilde{V}(q) =\frac{e^2m^2}{\vert q\vert^2(\vert q\vert^2-2m^2)}\\
\\
V(r)= \frac{e^2}{4 \pi}\left(\frac{cos(\sqrt{2}m r)}{r} \right)\\
\end{array}
\end{equation}

The Coulomb potential of the form of equation (\ref{vr}) forces a change in the hydrogen atoms' energy levels. There is a perturbation in the hydrogen atoms' Hamiltonian as follows:

\begin{equation}
\begin{array}{c}
V_0 = \frac{e^2}{4 \pi r},\\
\\
V=V_0+\Delta V= \frac{e^2}{4 \pi}\left(\frac{cos(\sqrt{2}m r)}{r} \right),\\
\\
\Delta H=\Delta V= \frac{e^2}{4 \pi} \frac{\left( cos(\sqrt{2}m r)-1\right)}{r}.\\
\end{array}
\end{equation}

where $V_0$ is the Coulomb potential, and $\Delta V$ is the perturbation term. The first-order perturbation for some energy levels are listed below (for $m= \frac{3.38}{R}, R= 46 (G ly)^{-1}$):

\begin{equation}
\Delta E= \langle\psi\vert \Delta V \vert  \psi \rangle
\end{equation}

Where $\psi = R_{nl}(r) Y_{lm}(\theta , \phi)$. For example:

\begin{equation}
\begin{array}{c}
R_{nl}=R_{10}\;\;\Rightarrow\; \Delta E=- O(10^{-25}) eV\\
\\
R_{nl}=R_{20}\;\;\Rightarrow\; \Delta E=- O(10^{-26}) eV\\
\\
R_{nl}=R_{21}\;\;\Rightarrow\; \Delta E=0. eV\\
\end{array}
\end{equation}

One can easily check that all energy levels have a miniscule non-measurable redshift.

\section{Conclusion}
\label{Co}

The goal of this manuscript was to investigate the effects of a multiverse system on the Green function of the Yang-Mills theory within the framework of non-commutative geometry. To research an observable evidence of the multiverse in our 4-D universe, we constructed a two-layer space. Here, each space corresponded to a universe and obtained the photon propagator in every single layer. Based on our calculations, the resultant photon spatial correlation function was comparable with the CMB and the X-ray two-point angular correlation. Our model predicts that all monochromatic wave two-point angular correlation has the same behaviour with the CMB. Based on our calculations and by fitting zeroes of our model and observed correlation curves, if there exists such a multiverse, the distance between the two layers lies within the order of the observable universe. To investigate more effects of the multiverse, the Coulomb potential and its effect on Hydrogen atom energy levels was re-derived by considering new photon propagator. It was observed that the perturbation of Hydrogen atom Hamiltonian and changes in energy levels are in the order of $10^{-25} eV$ and are so tiny that can be observed.

\appendix

\section{One-form and Curvature}
\label{A}

The most general form of Dirac operator is according to equation (\ref{dirac}):

\begin{equation}
D= \begin{pmatrix}
i \gamma_\mu (\partial^\mu+ \cdots) & \psi+ \gamma^5 m\\
\psi+ \gamma^5 m & i \gamma_\nu (\partial^\nu+ \cdots)\\
\end{pmatrix}.
\end{equation}

All of the zero order components of $D$ are denoted by dots. They do not contribute to the one-form presentation. (Algebra elements and zero order components are commutative.)
By using equation (2), one-form is shown to be in the following form:

\begin{equation}
    \omega= \sum_j a_0^j [D,a_1^j]= 
    \begin{pmatrix}
    \gamma^\mu \omega_{1\mu} & (\psi + \gamma^5 m) \phi\\
    -(\psi + \gamma^5 m) \tilde{\phi} & \gamma^\mu \omega_{2\mu}\\
    \end{pmatrix},
\end{equation}

where we use variables $a_{0}^j=a_{01}^j\oplus a_{02}^j$ and $a_{1}^j=a_{11}^j\oplus a_{12}^j$ then we have the following notations:

\begin{equation}
\begin{array}{cc}
    \omega_{1\mu} = \sum_j  a_{j1}^0 \partial_\mu a_{j1}^1 ; &     \phi= \sum_j a_{j1}^0 (a_{j2}^1-a_{j1}^1) ;\\
    \\
    \omega_{2\mu}= \sum_j a_{j2}^0 \partial_\mu a_{j2}^1 ; &    \tilde{\phi}= \sum_j a_{j2}^0 (a_{j2}^1-a_{j1}^1) .\\
\end{array}
\end{equation}

In order to have a self-adjoint one-form which can play the role of connection and vector potential the variables are chosen such that $\tilde{\phi} = \phi ^\dagger$.

Now the curvature $\theta$ can be derived.

\begin{equation}
\theta = \omega^2 + d \omega=\omega^2+[D, \omega]
\end{equation}

\begin{equation}
\omega^2 =\begin{pmatrix}
\gamma ^\mu \omega_{1 \mu} \gamma ^\rho \omega_{1 \rho}-\bar{\gamma}^2(\phi^{\dagger}\phi)  & (\phi) \gamma^\mu \gamma^5 (\omega_{1 \mu}- \omega_{2 \mu}) \\
-(\phi) \gamma^\mu \gamma^5 (\omega_{1 \mu}- \omega_{2 \mu})   &\gamma ^\nu \omega_{2 \nu} \gamma ^\rho \omega_{2 \rho}-\bar{\gamma}^2(\phi^{\dagger}\phi) \\
\end{pmatrix}
\end{equation}

\begin{equation}
\Scale[0.9]{
 d \omega =\begin{pmatrix}
\gamma^{\mu\nu}(\partial_\mu \omega_{1 \nu}-\partial_\nu \omega_{1 \mu})+2m\gamma^5(\phi-\phi^{\dagger}) & (m) \gamma^\mu \gamma^5 (\omega_{1 \mu} - \omega_{2 \mu})- m \gamma^{5} \gamma^{\mu} D_{\mu} \phi^\dagger \\
-(m) \gamma^\mu \gamma^5 (\omega_{1 \mu} - \omega_{2 \mu})+ m \gamma^{5} \gamma^{\mu} D_{\mu} \phi &\gamma^{\mu\nu}(\partial_\mu \omega_{2 \nu}-\partial_\nu \omega_{2 \mu})+2 m \gamma^5 (\phi-\phi^{\dagger}) \\
\end{pmatrix}}
\end{equation}

Notice that the first term of $\omega^2$ diagonal components are traceless so it does not contribute to the Lagrangian (\ref{Lagr}) and we omitted it in equation (\ref{scale}).

\section{Photon Propagator}
\label{B}

The Lagrangian of QED in the two-layer multiverse system is in the form of equation (\ref{Lag}). One may find an equation of motion by using the Euler-Lagrange equation, which results:

\begin{equation}
\begin{array}{c}
(\partial^2 g^{\mu\nu} -\partial ^\mu \partial^\nu) \mathbb{A}_{1\nu}=-m^2 (\mathbb{A}_1^\mu-\mathbb{A}_2^\mu)\\
\\
(\partial^2 g^{\rho\sigma} -\partial ^\rho \partial^\sigma) \mathbb{A}_{2\rho}=-m^2 (\mathbb{A}_2^\sigma-\mathbb{A}_1^\sigma)\\
\end{array}
\end{equation}

It is now possible to solve these two coupled equations:

\begin{equation}
\mathbb{A}_{2\mu}=\frac{1}{m^2}(\partial^2 g^{\nu \mu}-\partial ^\mu \partial ^\nu -m^2 g^{\nu \mu})\mathbb{A}_{1\nu}\\
\end{equation}

By substituting $\mu$ index with $\rho$ in the second line, we have the following relations:

\begin{equation}
\begin{array}{c}
\label{f}
\frac{\left( (\partial^2-m^2)g^{\mu\sigma}-\partial^\mu \partial^\sigma \right) \left( (\partial^2-m^2)g^\nu_\mu-\partial_\mu \partial^ \nu\right)\mathbb{A}_{1\nu}}{m^2}=m^2\mathbb{A}^{1\sigma}\\

\\

 \frac{\left(\left((\partial^2-m^2)g^{\mu\sigma}-\partial^\mu \partial^\sigma \right) \left( (\partial^2-m^2)g^\nu_\mu-\partial_\mu \partial^ \nu\right)-m^4 g^{\sigma\nu} \right) \mathbb{A}_{1\nu}}{m^2}=0\\
\end{array}
\end{equation}

The Fourier transform of the equation (\ref{f}) leads to the following relation:

\begin{equation}
\left(\frac{k^2+2m^2}{m^2}\right) \left( k^2 g^{\sigma\nu}-k^\sigma k^\nu \right) \tilde{D}_{F\;\nu\rho}=i\delta^\sigma_\rho
\end{equation}

One cannot derive the Feynman propagator from above equation because the coefficient of $\tilde{D}_{F\;\nu\rho}$ is a singular $4\times 4$ matrix. This problem occurs due to gauge symmetry. By using the Faddeev-Popov method \cite{Peskin}, the Lagrangian and equation of motions change, respectively, as shown in the following form:

\begin{equation}
\mathcal{L} \longrightarrow \mathcal{L}-\frac{1}{2\xi_1}(\partial^\mu \mathbb{A}_{1 \mu})^2-\frac{1}{2\xi_2}(\partial^\nu \mathbb{A}_{2 \nu})^2
\end{equation}

\begin{equation}
\begin{array}{c}
(\partial^2 g^{\mu\nu} -(1-\frac{1}{\xi_1})\partial ^\mu \partial^\nu) \mathbb{A}_{1\nu}=-m^2 (\mathbb{A}_1^\mu-\mathbb{A}_2^\mu)\\
\\
(\partial^2 g^{\rho\sigma} -(1-\frac{1}{\xi_2})\partial ^\rho \partial^\sigma) \mathbb{A}_{2\rho}=-m^2 (\mathbb{A}_2^\sigma-\mathbb{A}_1^\sigma)\\
\end{array}
\end{equation}

Specific values $\xi_1=1$ and $\xi_2=1$ are chosen to make the computation. Then, the result is a propagator in the form of an equation (\ref{prop}):

\begin{equation}
\tilde{D}_F^{\mu\nu}(k) = \frac{i m^2g^{\mu\nu}}{k^2(k^2+2m^2)- i \varepsilon}
\end{equation}

\section{Coulomb Potential}
\label{C}

Suppose scattering of two \textit{distinguishable} fermions. The leading order of contribution is then an equation (\ref{lead}):

\begin{equation}
\begin{array}{c}
i \mathcal{M}= -2 i e^2 \overline{u}(p') \gamma_\mu u(p) \frac{ g^{\mu\nu}  m^2}{q^2\left( q^2 +2m^2\right)-i \varepsilon}\overline{u}(k')\gamma_\nu  u(k)\\
\\
q=p'-p=k'-k\\
\end{array}
\end{equation}

To obtain the potential, in a non-relativistic limit:

\begin{equation}
\overline{u}(p') \gamma^0 u(p)= u^\dagger (p') u(p) \approx 2m \xi '^\dagger \xi
\end{equation}

The other terms,$\overline{u}(p') \gamma^i u(p)$ for $i=1,2,3$, can be neglected compare to $\overline{u}(p') \gamma^0 u(p)$ in non relativistic limit \cite{Peskin}.

Thus we have:

\begin{equation}
i \mathcal{M}= \frac{-2 i e^2 m^2}{\vert q\vert^2(\vert q\vert^2-2m^2)-i\varepsilon} (2m \xi '^\dagger \xi) (2m \xi '^\dagger \xi)
\end{equation}

We compare this with the Born approximation in which the scattering amplitude in non-relativistic quantum mechanics appears:

\begin{equation}
\langle p'\vert iT\vert p\rangle = -i\tilde{V}(q) (2\pi)\delta (E_{p'}- E_p)
\end{equation}

\begin{equation}
\tilde{V}(q) =\frac{2e^2m^2}{\vert q\vert^2(\vert q\vert^2-2m^2)-i\varepsilon}
\end{equation}

The repulsive potential can be derived from inversing the Fourier transform.

\begin{equation}
\begin{array}{c}
V(x)=\int \frac{\mathrm{d}^3 q}{(2\pi)^3} \frac{2e^2m^2}{\vert q\vert^2(\vert q\vert^2-2m^2)-i\varepsilon} \mathrm{e}^{iq.x}\\
\\
=\frac{2 e^2 m^2}{4 \pi ^2} \int_0^\infty \mathrm{d}\vert q \vert \;\left(\frac{\mathrm{e}^{i\vert q \vert r}-\mathrm{e}^{-i\vert q \vert r}}{i \vert q \vert r}\right) \frac{1}{\vert q \vert^2 -2m^2 -i \varepsilon}\\
\\
=\frac{2 e^2 m^2}{4 \pi ^2 r} \int_{-\infty}^\infty  \frac{\mathrm{e}^{i \vert q \vert r}}{\vert q \vert (\vert q \vert^2 -2m^2-i \varepsilon)}\\
\end{array}
\end{equation}

The contour of this integral can be closed above in the complex plane, and we pick up the residue of $0$ and $\sqrt{2 m^2+i \varepsilon }$. Therefore, we find the potential as follows:

\begin{equation}
V(r)= \frac{e^2}{4 \pi}\left( -\frac{1}{r}+\frac{2 cos(\sqrt{2}m r)}{r} \right)
\end{equation}


\end{document}